\documentclass [12pt]{article}
\makeatletter
\newenvironment{figurehere}
{\def\@captype{figure}} {} \makeatother
\usepackage{graphicx}
\usepackage{amsmath}
\usepackage{cite}
\begin{document}
\title{Arnold Tongues and Feigenbaum Exponents of the  Rational Mapping for $Q$-State Potts Model on Recursive Lattice: $Q<2$}

\author{L. N. Ananikyan$^1$, N. S. Ananikian$^{1}$, L. A. Chakhmakhchyan$^2$ \\[1mm]
{\small \sl $^1$Yerevan Physics Institute, Alikhanian Br. 2,
0036 Yerevan, Armenia,} \\
{\small \sl $^2$ Yerevan State University, A. Manoogian 1,
Yerevan 0025, Armenia}\\[1mm]}
\date{}

\maketitle
\thispagestyle{empty}
\begin{abstract}
We considered $Q$-state Potts model on Bethe lattice in presence of
external magnetic field  for $Q<2$ by means of recursion relation
technique. This allows to study the phase transition mechanism in
terms of the obtained one dimensional rational mapping. The
convergence of Feigenabaum $\alpha$ and $\delta$ exponents for the aforementioned
mapping is investigated for the period doubling and three cyclic
window. We regarded the Lyapunov exponent as an order parameter for
the characterization of the model and discussed its dependence on
temperature and magnetic field. Arnold tongues analogs with winding numbers
$\mathrm{w}=1/2$, $\mathrm{w}=2/4$ and $\mathrm{w}=1/3$ (in the
three cyclic window) are constructed for $Q<2$. The critical
temperatures of the model are discussed and their dependence on $Q$
is investigated.  We also proposed an approximate method for
constructing Arnold tongues via Feigenbaum $\delta$  exponent.
\vspace{10pt}

\noindent\textit{Keywords}: Bifurcation; Feigenbaum exponents;
Universality; Modulated phases; Arnold tongues
\end{abstract}

\twocolumn
\section{Introduction}

The integer $Q$-state Potts model is a generalization
of the Ising model to more-than-two components. Some particular
cases of the model (different integer values of $Q$) played an
essential role in the theory of phase transitions and critical
phenomena \cite{Baxter, phase}. For $Q\rightarrow1$ it is
associated with the percolation (connectivity) model \cite{Hamm,
percolation}. Both dilute \cite{Dilute} and "pure" Potts models can
be also represented by Kasteleyn-Fortuin (KF) clusters
\cite{Fourtin2}. The partition function of the Potts model is well
defined for any non-integer $Q\geq0$ \cite{noninteger}. Some
interesting special cases arise in the limit $Q\rightarrow0$: e.g.
the limit $Q;v=e^{\beta J}-1\rightarrow0$ with $u=\frac{v}{Q}$ held
fixed gives rise to a model of spanning forest \cite{Jacobsen,
Spanning}.

Potts model can be also realized in experiments. In particular,
$Q$-state Potts lattice model can be used for description of cold
denaturation of proteins in solvents \cite{Anfinsen, Salvi}.
Research of experimental realizations of the model shows that in the
case $0<Q<1$ it is in the same universality class of the transition
in gelation and vulcanization processes in branched polymers
\cite{Lubensky}. Besides, many physical processes like the resistor
network, dilute spin glass, self organizing critical systems can be
formulated in terms of the model, when $Q<2$ \cite{Duplantier,
Whittle, Aharony}.

Since the Potts model generally is not exactly solved (there isn't
any solution of the model in dimensions $d>2$ in non-zero magnetic
field), approximate methods are required. An important advance was
made with the invention of the Swendsen-Wang (SW) and \\
Chayes-Machta (CM) \cite{Swendson}  algorithms for simulating the
ferromagnetic Potts model at positive integer $Q$ and the
random-cluster model with any real $Q\geq1$ respectively (for a
summary see \cite{SW}). No analytical results can be obtained in
these cases.

In this paper we used the Bethe approximation. One of the advantages
of this kind approach is that some models can be solved here exactly
with the help of the recurrence relation technique \cite{Ananikian}
and analytical results can be obtained. The method is more reliable
than mean-field  calculations \cite{Gujrati}. This approach can be
also applied to the generalized Bethe lattice (known as Husimi
lattice), which can be used for investigation of models with
multisite interactions \cite{Monroe}. It was used for the study of
magnetic properties of the solid $^3\mathrm{He}$ films \cite{He3}.
In the case of recurrence relation technique, the properties of the
model are associated with the behavior of the iteration sequence
$\{x_n\}$ of a nonlinear mapping, which mostly exhibits period
doubling cascade, chaos, $p$-cyclic windows. Here the Lyapunov
exponent serves not only a good order parameter, but also gives
relevant information about the geometrical and dynamical properties
of the model's attractors \cite{Lyapunov}. The phase structure of
the model can be investigated with the help of mapping's Arnold
tongues \cite{Arnold}.

The concepts Scaling and Universality have played an essential
role in the description of statistical systems
\cite{Kadanoff}. Particularly the behavior of
famous Feigenbaum exponents \cite{Feigenbaum2, Feigenbaum3}
for one dimensional rational mapping describing statistical model
on Husimi lattice was investigated  \cite{Lusinyants}.

This paper is organized as follows. In the next section we present
the recursion relation and phase structure of the Potts model on
the Bethe lattice. In Sec.~\ref{feig} the universality of
Feigenbaum exponents for the obtained one-dimensional rational
mapping is derived (for $Q<2$). The dependence of the Lyapunov exponent on external magnetic field and
temperature is given in Sec.~\ref{lyap}.  In Sec.~\ref{arnold}
Arnold tongues analogs are constructed for $Q<2$. In Sec.~\ref{three} we investigated the three periodic window of the mentioned above mapping. The critical temperatures and an approximate me\-thod for constructing Arnold tongues through
Feigenbaum $\delta$ exponent are introduced in Sec.~\ref{arnold1}. Finally we summarize our
results in Sec.~\ref{concl}.

\section{Rational Mapping for the Potts Model and Phase Structure\label{map}}

The $Q$-state Potts model in the case of two-site
interactions in presence of external magnetic field is defined by
the Hamiltonian
\begin{eqnarray}
{\mathcal{H}}=-J\sum_{(i,j)}\delta({{\sigma_{i},\sigma_{j}} })-H
\sum_{i}\delta({{\sigma_{i},Q} }). \label{1}
\end{eqnarray}
The first sum in Eq.~(\ref{1}) goes over all nearest-neighbor pairs
and the second one over all sites of the lattice ($J<0$ is the anti-ferromagnetic and $J>0$ the ferromagnetic case).
The partition function and single site magnetization is given by
\begin{gather}
\mathcal{Z}=\sum_{\{{\sigma}\}}{e^{-\frac{\mathcal{H}}{k_B T}}},\label{2}
\\ M=\langle\delta(\sigma _0,Q)\rangle=\mathcal{Z}^{-1}\sum_{\{\sigma\}}{\delta(\sigma _0,Q)}e^{-\frac{\mathcal{H}}{k_BT}}, \label{3}
\end{gather}
where $k_B$ is the Boltzmann constant (we will set $k_B=1$). The Bethe lattice can be separated into $\gamma$ identical
branches by cutting apart at the central point. The partition
function can be written as
\begin{eqnarray}
\mathcal{Z}_{n}=\sum_{\{{\sigma} _0\}}exp\{\frac{H}{T}\cdot\delta(\sigma _0,
Q)\}[g_{n}(\sigma _0)]^\gamma, \label{4}
\end{eqnarray}
where $\sigma_0$ is the central spin and $g_{n}(\sigma _0)$ is the
contribution of each lattice branch. Following well-known procedure \cite{Monroe, Ananikian, He3}, we obtain
\begin{gather}
x_n=f(x_{n-1}), \nonumber
\\ f(x)=\frac{e^{\frac{H}{T}}+(e^{\frac{J}{T}}+Q-2)x^{\gamma-1}}{e^{\frac{H+J}{T}}+(Q-1)x^{\gamma-1}}, \label{7}
\end{gather}
where
\begin{eqnarray}
x_n=\frac{g_n(\sigma\neq Q)}{g_n(\sigma=Q)}.\label{6}
\end{eqnarray}
Equation (\ref{7}) is called Potts-Bethe mapping. Knowing recursion
relation we can calculate the magnetization of the central spin:
\begin{eqnarray}
M_n=\langle\delta(\sigma_0, Q)\rangle
=\frac{e^\frac{H}{T}}{e^\frac{H}{T}+(Q-1)x_{n}^\gamma}.\label{8}
\end{eqnarray}

\begin{center}
\begin{figure*}
\begin{tabular}{ c c }
\small{(a)} & \small{(b)}\\
\begin{figurehere}
\includegraphics[width=7cm]{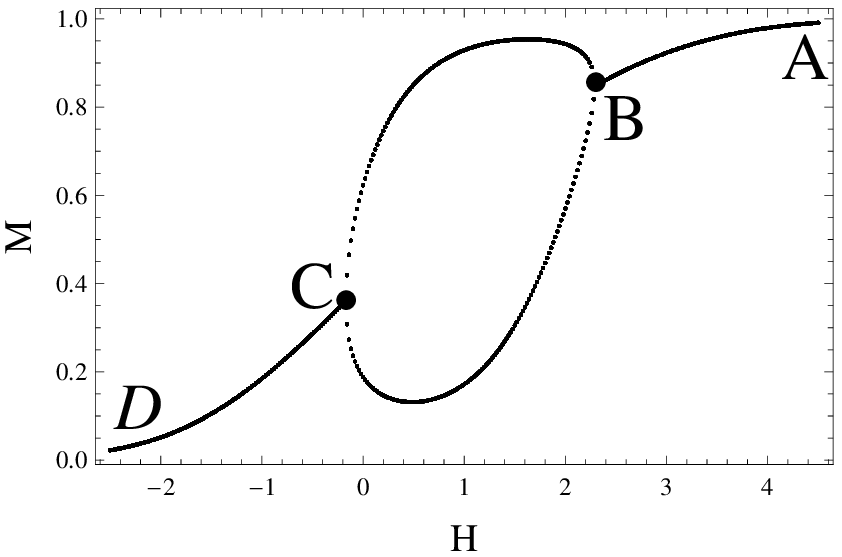}
\end{figurehere} &  \begin{figurehere}
\includegraphics[width=7cm]{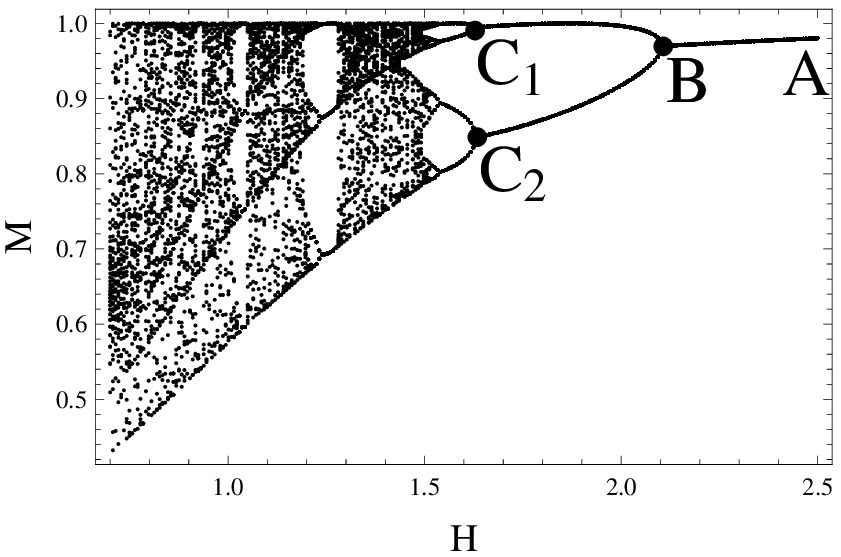}
\end{figurehere}
\end{tabular}
\caption{ \small{The plot of $M$ (magnetization) versus external
magnetic field $H$ for $\gamma=3$, (a) $Q=2.5$, $J=-1$, $T=0.5$ (the areas $AB$ and $CD$ correspond to phase without sub-lattice structure, $BC$ to antiferromagnetic phase, $B$ and $C$ are the points at which the phase transition between these phases takes place); (b) $Q=1.1$, $J=-1$, $T=0.5$ ($AB$ correspond to the magnetization of the paramagnteic phase, $BC_1$ and $BC_2$ to magnetization of two sub-lattices in the antiferromagnetic phase, $B$ is the point of phase transition between paramagnetic and antiferromagnetic phases, $C_1$ and $C_2$ the points of transition between antiferromagnetic and four-periodic modulated phases).
\label{fig2}}}
\end{figure*}
\end{center}

In real statistical systems the bifurcation points of mappings like
Eq.~(\ref{7}) correspond to phase transition points of second order.
For systems with $Q<2$ and anti-ferromagnetic coupling ($J<0$), the
dependence of $M$ magnetization on the field $H$ is complicated:
full range of period doubling, chaos, $p$-cyclic windows can be
observed here. Figure~\ref{fig2} shows bifurcation diagrams for
different values of parameters.

In terms of the mapping in Eq.~(\ref{7}) the area, where $M$ is
single-valued function of $H$ [$AB$ and $CD$ in Fig.~\ref{fig2}(a),
$AB$ in Fig.~\ref{fig2}(b)], the recursion sequence $\{x_n\}$ of the
mapping converges to one stable fixed point $x_0$ (phase without
sub-lattice structure). The area after first bifurcation point
[$BC$ in Fig.~\ref{fig2}(a), $C_1BC_2$ in
Fig.~\ref{fig2}(b)], sequence $\{x_n\}$ converges to two stable
points. Therefore here we have two values of magnetization, which
should be explained, as an arising of two sub-lattices of
anti-ferromagnetic order. The areas between consequent bifurcation
points are described by a sequence of $n$ fixed points (existence of
$n$ sub-lattices), corresponding to various modulated phases with
finite period (commensurate modulated phases), which also appear
in $p$-cyclic windows.

\section{Universality of \\ Feigenbaum Exponents \\ for Potts-Bethe Mapping\label{feig}}

Let $r_n$ be the value of the parameter $r$ of a mapping at the period doubling bifurcation point,
$r_\infty$ the value of $r$
from which the chaotic behavior ensues. It turns out that the
values of $r_n$ satisfy the following scaling:
\begin{equation}
r_n=r_{\infty}-const\delta^{-n}, \qquad   n\gg1.  \label{23}
\end{equation}

If $d_n$  is the distance between the point $x^*$
for which  $f_{r}(x)$ is extremal and the nearest point on the $2^n$ cycle
(Fig.~\ref{fig9}), then
\begin{equation}
\frac {d_n}{d_{n+1}}=-\alpha, \qquad   n\gg1.    \label{24}
\end{equation}

The quantities $\alpha$ in Eq.~(\ref{23}) and $\delta$ in Eq.~(\ref{24}) are
Feigenbaum exponents \cite{Feigenbaum3, Feigenbaum2}:
\begin{subequations}
\begin{align}\label{25b}
\delta=4,6692016091...\\ \label{25a} \alpha=2,50290787050...
\end{align}
\end{subequations}
If $R_n$ is the value of the parameter at which the
line $x=x^{*}$ intersects $2^{n}$ periodic cycle
(Fig.~\ref{fig9}), then
\begin{equation}
R_n=R_{\infty}-const'\delta^{-n}. \label{26.1}
\end{equation}
The exponents $\delta$ and $\alpha$ are universal, \textit{i.e.} \\
Eqs.~(\ref{23}), (\ref{24}) and (\ref{26.1}) are true for wide
variety of mappings, $\alpha$ and $\delta$ having the same values
as in Eqs.~(\ref{25b}) and (\ref{25a}) \cite{Feigenbaum3}. For the $const$ and $const'$, presented in Eqs.~(\ref{23}) and
(\ref{26.1}), they depend on family of reflection functions.

Introducing $\delta_n$
\begin{equation}
\delta_{n}=\frac{R_n-R_{n-1}}{R_{n+1}-R_{n}}, \label{26}
\end{equation}
taking into account that \cite{Feigenbaum2}
\begin{equation}
\delta=\lim_{n\rightarrow\infty}\delta_n \label{27}
\end{equation}
and the fact that
\begin{equation}
f^{(2^{n})}_{R_n}(x^{*})=x^{*}\label{28}
\end{equation}
($f^{(n)}(x)$ is the $n-th$ iteration of $f(x)$, \textit{i.e.}
$f^{(2)}(x)=f[f(x)]$, $f^{(2^2)}(x)=f^{(4)}(x)=f^{(2)}[f^{(2)}(x)]$, etc.),
we can calculate Feigenbaum exponents for a mapping $f(x)$.
In our case
\begin{eqnarray}
\begin{split}
f(x)&=\frac{e^{\frac{H}{T}}+(e^{\frac{J}{T}}+Q-2)x^{\gamma-1}}{e^{\frac{H}{T}}\cdot
e^{\frac{J}{T}}+(Q-1)x^{\gamma-1}}
\\&=\frac{r+(e^{\frac{J}{T}}+Q-2)x^{\gamma-1}}{r\cdot
e^{\frac{J}{T}}+(Q-1)x^{\gamma-1}}, \label{32}
\end{split}
\end{eqnarray}
where
\begin{equation}
r=e^{\frac{H}{T}}.\label{31}
\end{equation}
\begin{figurehere}
\begin{center}
\includegraphics[width=8cm]{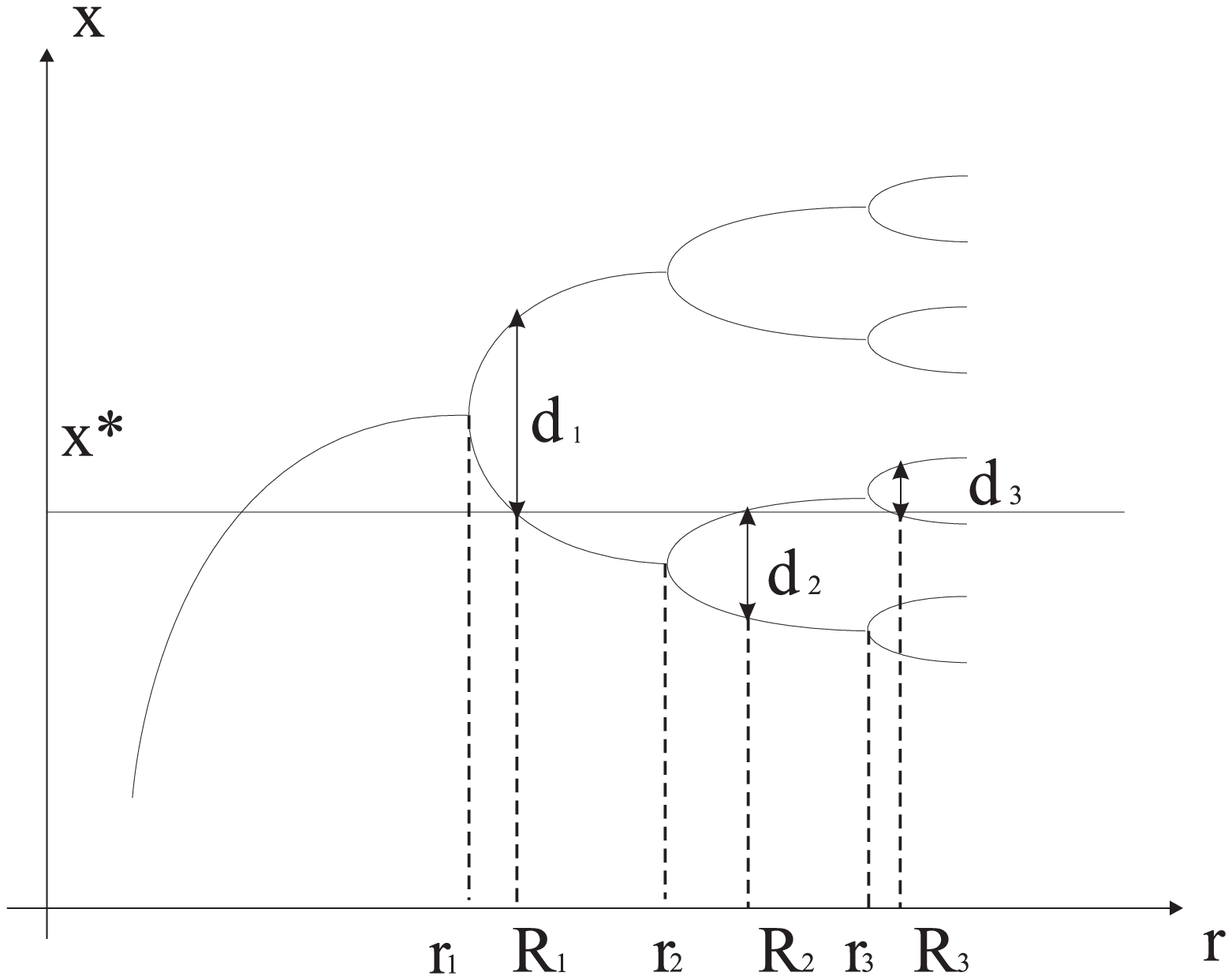}
\caption { \small { The period doubling process (schematically).}}
\label{fig9}
\end{center}
\end{figurehere}
Mapping $f(x)$ is extremal for $x^*=0$.

The values of $R_n$ and $r_n$ must satisfy following condition (see Fig.~\ref{fig9}):
\begin{equation}
r_1<R_1<r_2<R_2<r_3<... \label{31.1}
\end{equation}
We start solving Eq. (\ref{28}) from $n=0$,
putting values of $R_n$ in order (\ref{31.1}). Besides, one can see, that
\begin{equation}
\lim_{n\rightarrow\infty}R_n=R_\infty=r_\infty. \label{31.2}
\end{equation}

Our numerical calculations for $J=-1$,
$T=1$,$\gamma=3$, $Q=1.1$ and $Q=0.8$ are shown in Table~\ref{table}.

We find that values of $\alpha$ and $\delta$  converge to the Feigenbaum exponents for the Potts-Bethe
mapping, which describes physical properties of a real statistical systems. The convergence in the case $0<Q<1$ is
faster than in the case $1<Q<2$ (see Table~\ref{table}).

For $const'$, presented in Eq.~(\ref{26.1}), ini the case $Q=0.8$ and $Q=1.1$
we obtained the following values: $const'=-2.682...$ and
$const'=-5.034...$ respectively.

\section{Lyapunov Exponents \\ for One Dimensional \\ Rational Mapping \label{lyap}}
It is well known that under action of the mapping
$x_{n+1}=f(x_n)$ two nearby points can be dispersed. The Lyapunov
exponent $\lambda(x)$ characterizes the degree of the
exponential divergence of two adjacent points. The exact formula for the Lyapunov
exponent is:
\begin{eqnarray}
\begin{split}
\lambda{(x)}=\lim_{n\rightarrow\infty}\lim_{\varepsilon\rightarrow0}{\frac{1}{n}}\ln&\left|\frac{f^{(n)}(x+\varepsilon)-
f^{(n)}(x)}{\varepsilon}\right|\\
&=\lim_{n\rightarrow\infty}{\frac{1}{n}}\ln\left|\frac{df^{(n)}(x)}{dx}\right|.\label{21}
\end{split}
\end{eqnarray}

\begin{center}
\begin{figure*}
\begin{tabular}{ c c }
\small(a) & \small(b)\\
 \begin{figurehere}
\includegraphics[width=7cm]{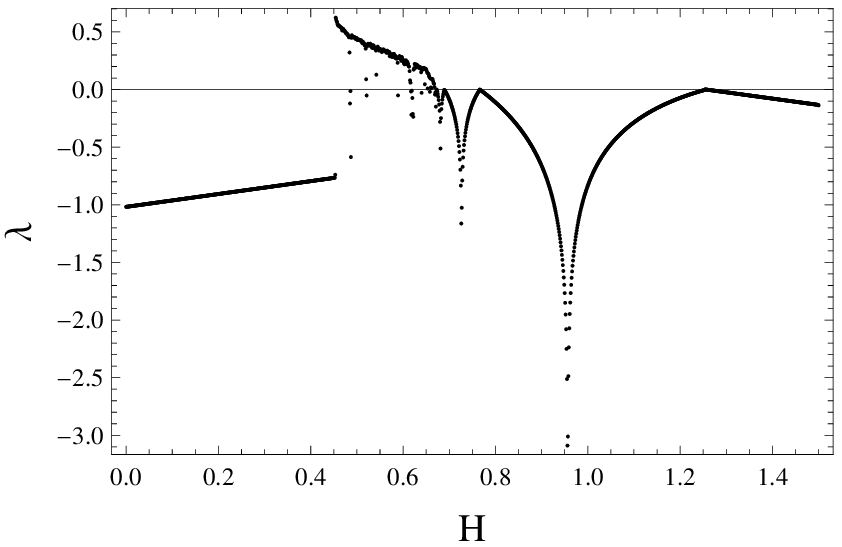}
\end{figurehere} &  \begin{figurehere}
\includegraphics[width=7cm]{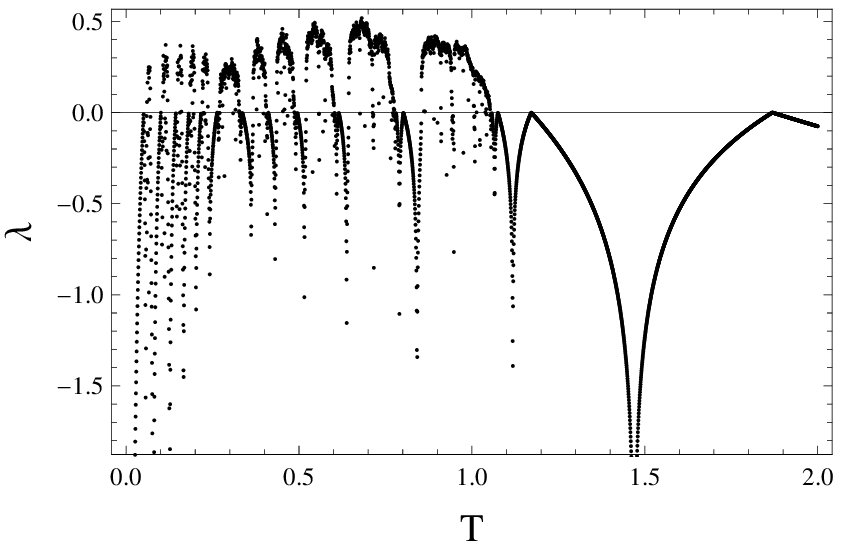}
\end{figurehere}  \\
\end{tabular}
\caption {\small{ The plot of the Lyapunov exponent for the
Potts-Bethe mapping [Eq.~(\ref{7})] (a) versus magnetic field $H$ for $Q=0.8$, $J=-1$,
$\gamma=3$, $T=2$; (b) versus temperature $T$ for $Q=1.2$, $J=-1$, $\gamma=3$, $H=0.2$. \label{Lyap}}}
\end{figure*}
\end{center}

We regard the dependence of
$\lambda(x)$ for the Potts-Bethe mapping [Eq.~(\ref{7})] on the external
magnetic field $H$, fixing $Q$, temperature $T$ and strength of
interaction $J$ [Fig.~\ref{Lyap}(a)].

One can also study the dependence of the Lyapunov exponent on the temperature $T$, fixing $H$, $Q$, and $J$ [Fig.~\ref{Lyap}(b)]. This will help to examine the behavior of the magnetization at fixed external magnetic field, when temperature is varied (see also Secs.~\ref{arnold} and \ref{arnold1}).

Figures~\ref{Lyap}(a) and (b) indicates, that at fixed $H$ the
chaotic region is richer than at fixed $T$: a large variety of
different $p$-cyclic windows can be observed in Fig.~\ref{Lyap}(b).

One of the major properties of the Lyapunov exponent is that it is
equal to zero at the bifurcation point. This property can be
easily obtained from the fact that the bifurcation point
corresponds to a neutral point of the mapping.

\section{Arnold Tongues\label{arnold}}

Values of parameters (external magnetic field and temperature) at
phase transition (bifurcation) points can be found as a neutral point of the mapping:
\begin{equation}
f'(x)=e^{i\varphi}. \label{28.6}
\end{equation}

\noindent Different values of $\varphi$ correspond to different
types of bifurcation:

\noindent1. $\varphi=2 \pi n$. This case corresponds to saddle-node or tangent
 bifurcation (type I intermittency) \cite{Pomeau1, Pomeau2}. It can be observed e.g.
 in the case of the logistic map as $p$-cyclic window in the chaotic regime (see also Sec.~\ref{three}).

\noindent 2. $\varphi=\pi +2\pi n$. Here we have bifurcation
corresponding to period doubling \cite{Schuster} (type
III intermittency).

\noindent3. A pair of conjugate complex eigenvalues of the Jacobian in the case of multidimensional mapping corresponds to the Hopf-bifurcation (type II intermittency) which introduces new basic frequencies in the system at the bifurcation points \cite{Hopf}.

At the first bifurcation point in period doubling regime we have:
\begin{eqnarray}
\left\{\begin{array}{ll}
f(x)=x & \\
f'(x)=-1. &
\end{array} \right.\label{33}
\end{eqnarray}
Eliminating $x$ from Eq.~(\ref{33}) for the mapping in Eq. (\ref{7}) and solving obtained equation in order to $H$, we obtain two branches:
\begin{eqnarray}
\begin{split}
   H&=T \cdot\ln \left[ \frac{1}{8(-1+Q)}e^{-\frac{3J}{T}}\left(-6 e^{\frac{3J}{T}}\right.\right.\\
   &\times(Q-2)-e^{J/T} (6Q+u-6) (Q-2)\\
   &-3 e^{\frac{4J}{T}}+(Q-1) (Q+u-1)\\
   &\left.\left.-e^{\frac{2J}{T}} (3(Q-2)Q+u+6\right)\right] \label{36}
\end{split}
\end{eqnarray}
and
\begin{eqnarray}
\begin{split}
   H&=T \cdot\ln \left[ \frac{1}{8(-1+Q)}e^{-\frac{3J}{T}}\left(-6 e^{\frac{3J}{T}}\right.\right.\\
   &\times(Q-2)-e^{J/T} (6Q-u-6) (Q-2)\\
   &-3 e^{\frac{4J}{T}}+(Q-1) (Q-u-1)\\
   &\left.\left.+e^{\frac{2J}{T}} (-3(Q-2)Q+u-6\right)\right],\label{37}
\end{split}
\end{eqnarray}
where
\begin{eqnarray}
\begin{split}
   u&=\sqrt{-1+e^{J/T}} \sqrt{Q+e^{J/T}-1} \\
   &\times\sqrt{9 e^{J/T} Q-Q-18 e^{J/T}+9e^{\frac{2J}{T}}+1}.\label{37.55}
\end{split}
\end{eqnarray}
Let us consider two cases:

\noindent 1. $Q>1$. In this particular case two branches in
Eqs.~(\ref{36}) and (\ref{37}) define the line separating
paramagnetic (P) and anti-ferromagnetic (2AF1, \textit{i.e.} $2^1=2$ periodic) phases.

\noindent 2. $Q<1$. This case is more complicated. The reason is
following (here and further $\gamma=3$): when $Q<1$, Potts-Bethe  mapping  has singular
points (see  Fig.~\ref{fig2.1}).
\begin{figurehere}
\begin{center}
\includegraphics[width=8cm]{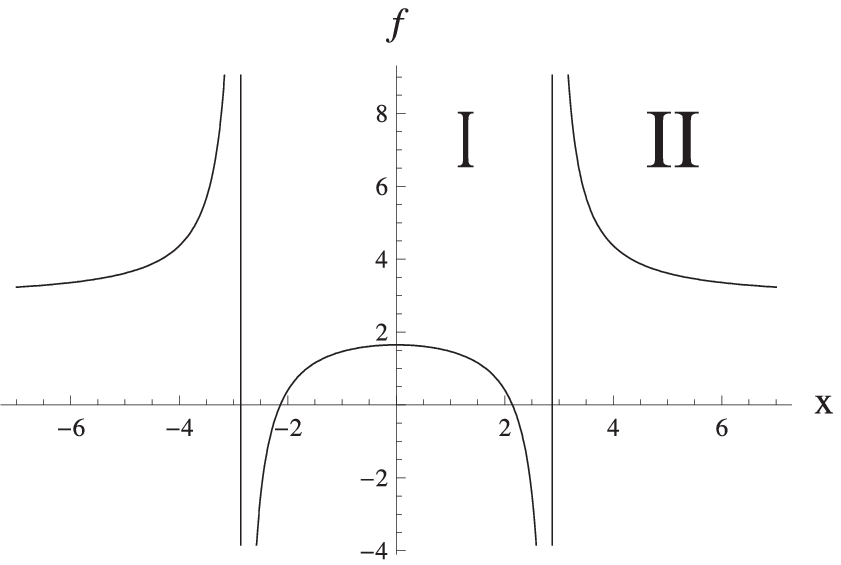}
\caption { \small { The plot of Potts-Bethe mapping $f(x)$ [Eq.~(\ref{7})] for the case
$Q=0.8$, $J=-1$, $\gamma=3$, $T=2$, $H=2$.}} \label{fig2.1}
\end{center}
\end{figurehere}
As one can see from Fig.~\ref{fig2.1}, the behavior of the mapping becomes
sensible to initial point of the iteration: stable point can fall
into area I or II .

\begin{figure*}
\begin{center}
\includegraphics[width=10cm]{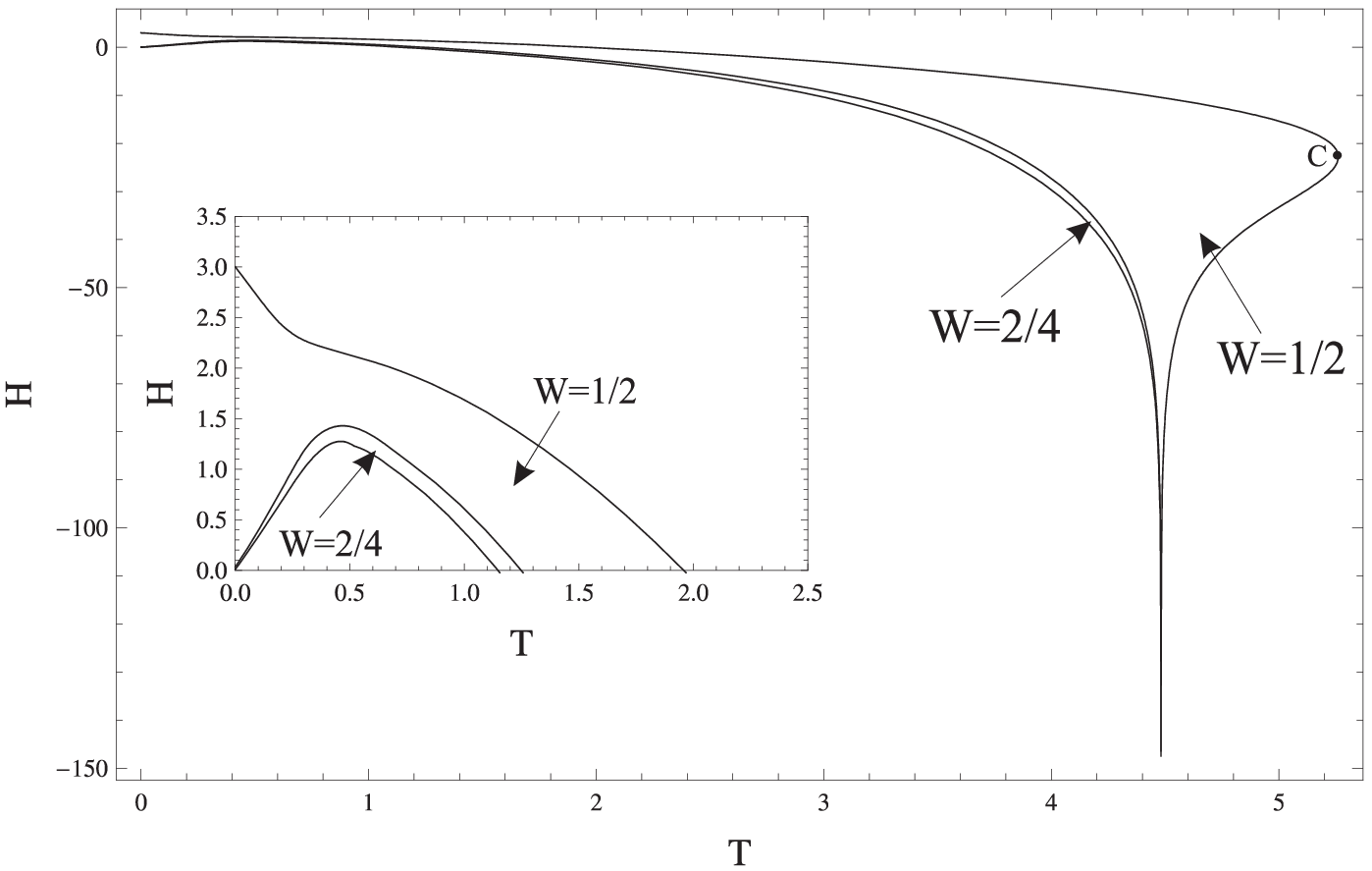}
\caption { \small { Arnold tongues with winding numbers
$\mathrm{w}=1/2$ and $\mathrm{w}=2/4$ for Potts model for
$Q=1.2$, $J=-1$, $\gamma=3$ (the insert shows details in the area
$H>0$). Point $C$ describes the upper bound of the temperature
($T_c=5.260340$).\label{ar1}}}
\end{center}
\end{figure*} 
But it turns out, that only area I has
physical meaning: in case $Q<1$,
magnetization can be negative, which has not physical
interpretation [see Eq.~(\ref{3})]. Since $x_n$ converges to stable $x_0$ point, therefore
the criterion $M$ being positive is
\begin{equation}
x_0 < \sqrt[3]{\frac{e^{\frac{H}{T}}}{1-Q}}. \label{37.8}
\end{equation}

One can easily show that in area II condition (\ref{37.8}) is violated, hence here $M<0$.
So, in our further investigations we will assume, that stable points
of the mapping are in area I. This case corresponds to the
Eq.~(\ref{37}).
It is also of interest to find the line separating
anti-ferromagnetic (2AF1) and four - periodic (2M2, \textit{i.e.} $2^2=4$ periodic) modulated phases.
This means to find the values of $T$ and $H$, where the second
bifurcation occurs. The procedure is the same with the only
difference being that here second iteration of the mapping
($f^{(2)}(x)=f[f(x)]$) looses its stability:
\begin{eqnarray}
\left\{\begin{array}{ll}
f^{(2)}(x)=x & \\
(f^{(2)}(x))'=-1. &
\end{array} \right.\label{37.2}
\end{eqnarray}
The area between two curves, obtained from Eqs.~(\ref{33}) and
(\ref{37.2}) (anti-ferromagnetic phase), is the Arnold tongue analog with winding number $\mathrm{w}=1/2$. Using the same technique we
can also find the line, separating four - periodic (2M2) and eight -
periodic (2M3, \textit{i.e.} $2^3=8$ periodic) modulated phases, \textit{i.e.} the values of $T$ and $H$,
where the third bifurcation occurs. The condition will be:
\begin{eqnarray}
\left\{\begin{array}{ll}
f^{(4)}(x)=x & \\
(f^{(4)}(x))'=-1. &
\end{array} \right.\label{37.9}
\end{eqnarray}
The area between curves defined by Eqs.~(\ref{37.2}) and
(\ref{37.9}) is the Arnold tongue analog with winding number
$\mathrm{w}=2/4$ (the area of existence four - periodic modulated
phase). The result is given in Figs.~\ref{ar1} and \ref{ar2}.

It was already mentioned in Sec.~\ref{map} that convergence of Feigenbaum exponents in areas $0<Q<1$ and  $1<Q<2$ are different. As one can see from Figs.~\ref{ar1} and \ref{ar2}, Arnold tongues are different in these areas, too.

\begin{figurehere}
\begin{center}
\includegraphics[width=7cm]{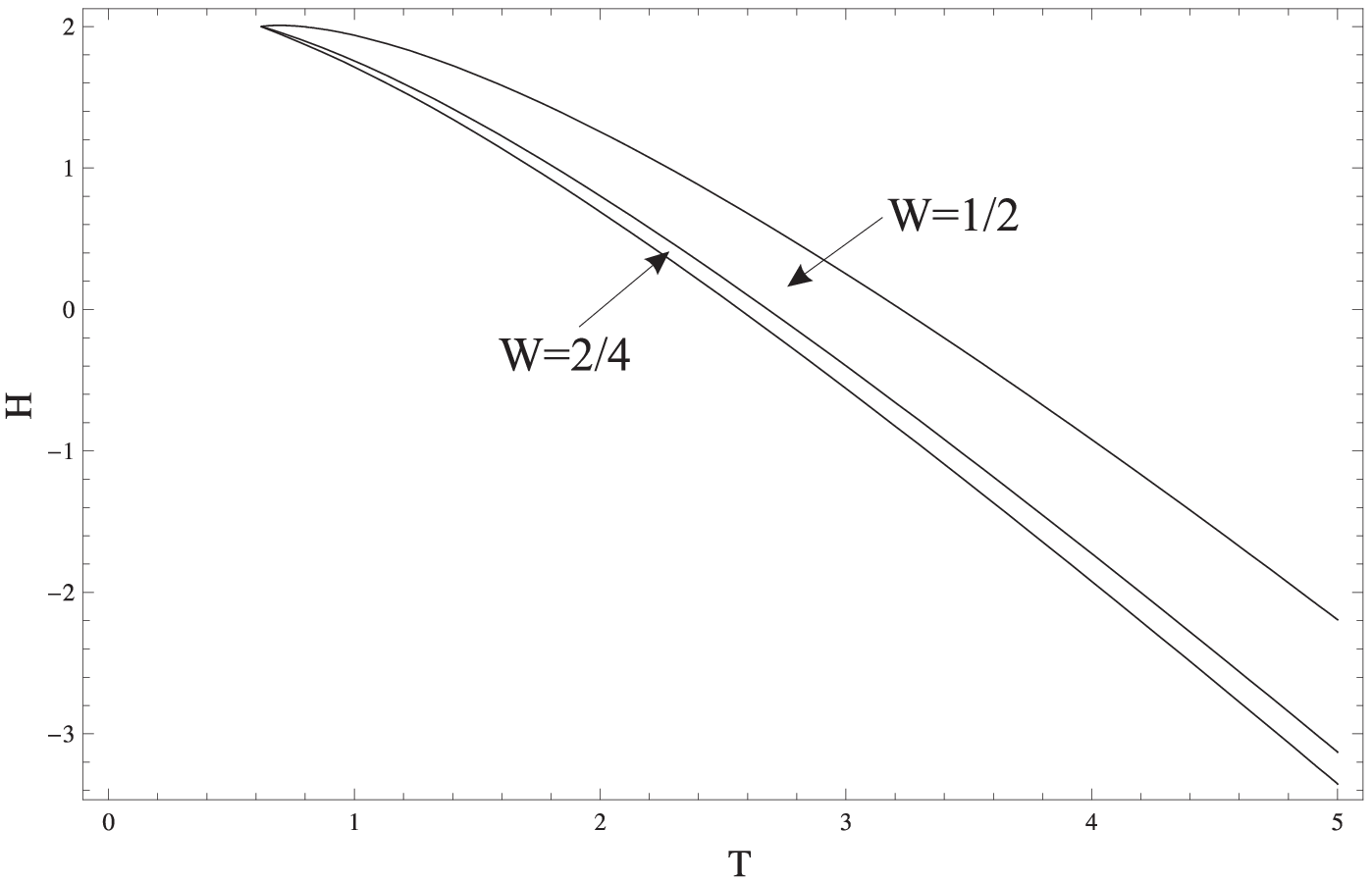}
\caption { \small { Arnold tongues with winding numbers
$\mathrm{w}=1/2$ and $\mathrm{w}=2/4$ for Potts model for $Q=0.8$, $J=-1$, $\gamma=3$. \label{ar2}}}
\end{center}
\end{figurehere}

\begin{center}
\begin{figure*}
\begin{tabular}{ c c }
\small{(a)} & \small{(b)}\\
\begin{figurehere}
\includegraphics[width=7cm]{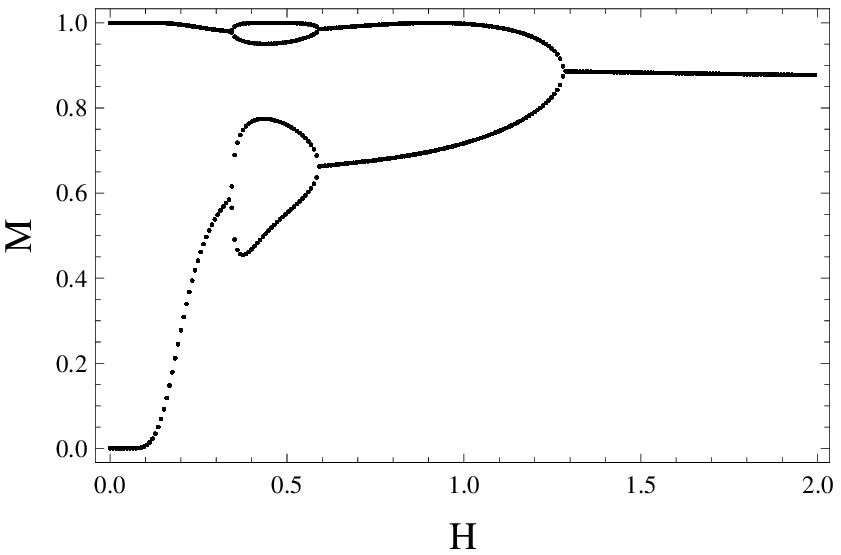}
\end{figurehere} & \begin{figurehere}
\includegraphics[width=7cm]{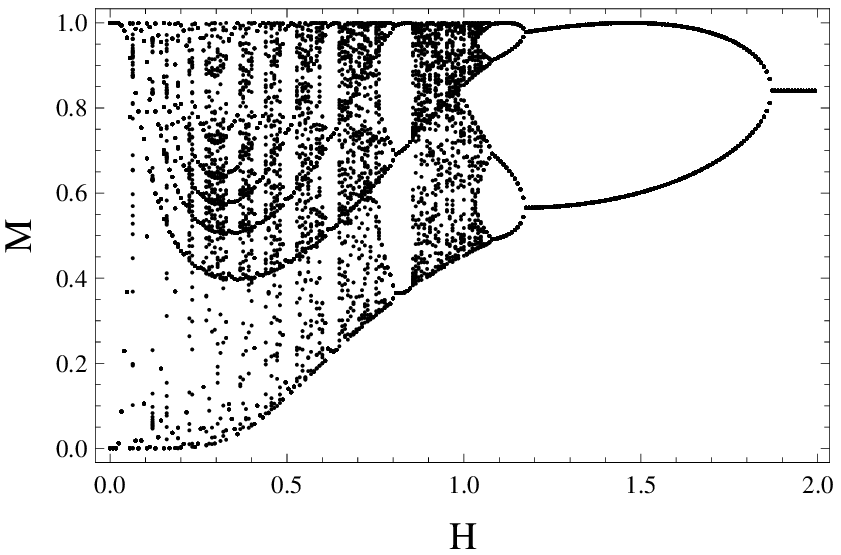}
\end{figurehere}\\
\end{tabular}
\caption { \small{The plot of $M$ (magnetization) versus
temperature $T$ for $Q=1.2$, $J=-1$, $\gamma=3$ (a) $H=1.3$; (b) $H=0.2$. \label{fig3.13}}}
\end{figure*}
\end{center}

$M$ magnetization can be regarded as a function of $T$, fixing the
value of external magnetic field. Fig.~\ref{ar1} indicates, that lowering  $H$  first two
periodic cycle appears, corresponding to P-2AF1 transition.
When the line $H$ intersects the upper curve of the Arnold tongue
with $\mathrm{w}=2/4$, one finds a bubble on each branch [Fig.~\ref{fig3.13}(a)]. Here we have P-2AF1-2M2-2AF1 transitions. After the line $H$ intersects the lower curve, we have another two bubbles
which points on P-2AF1-2M2-2M3-2M2-2AF1 transitions. With further
reduction of external magnetic field new bubbles are formed as
parts of the old ones and for still lower $H$'s we reach a region
of chaotic behavior [Fig.~\ref{fig3.13}(b)].

\section{Three Periodic Window \label{three}}

In this section we will investigate the three
periodic window of the Potts-Bethe mapping. Some definite values of
parameters $T$ and $H$ form a line in the parameter space, which
separates the chaotic and three periodic regimes.

\begin{figurehere}
\begin{center}
\includegraphics[width=7cm]{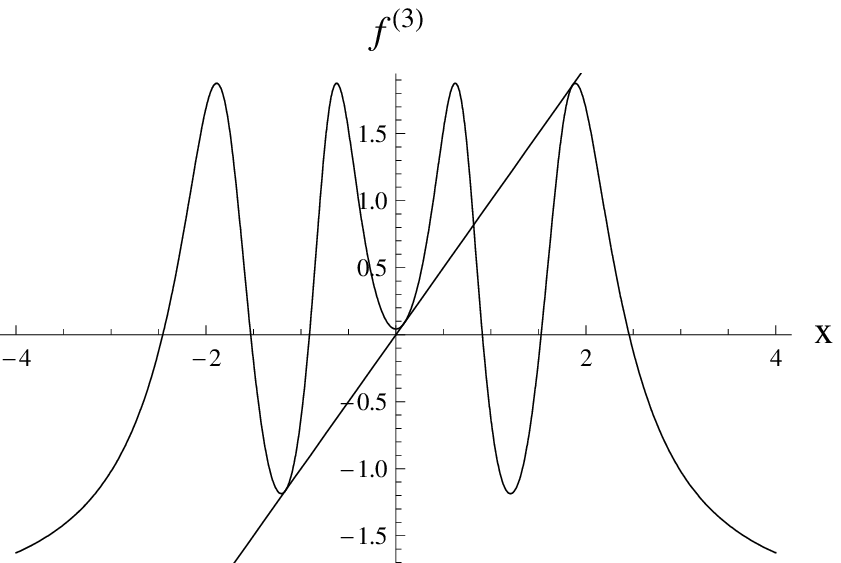}
\caption { \small { The Plot of $f^{(3)}(x)$ [third iteration of the Potts-Bethe mapping in Eq.~(\ref{7})] for $Q=1.1$, $J=-1$, $\gamma=3$ at one point of the saddle-node bifurcation line defined by Eq.~(\ref{50}): $T=1.5915$, $H=-1$. \label{3n}}} \end{center}
\end{figurehere}
Figure~\ref{3n}
presents the third iteration of the mapping $f(x)$ [Eq.~(\ref{7})]
for one point from that line (as one can see the mapping has three
fixed points here). On this line the saddle-node bifurcation takes
place, \textit{i.e.} the aforementioned line can be found from the
following condition (see Sec.~\ref{arnold}):
\begin{eqnarray}
\left\{\begin{array}{ll}
f^{(3)}(x)=x & \\
(f^{(3)}(x))'=1. &
\end{array} \right.\label{50}
\end{eqnarray}

Consequent bifurcations correspond to period doubling, which leads
to appearance of stable $3\ast2^n$ periodic cycles. Following the
technique described in Sec.~\ref{arnold}, we can find the line
separating three and six periodic cycles (three and six periodic
modulated phases) from the following system of equations:

\begin{eqnarray}
\left\{\begin{array}{ll}
f^{(3)}(x)=x & \\
(f^{(3)}(x))'=-1. &
\end{array} \right.\label{51}
\end{eqnarray}

\begin{figurehere}
\begin{center}
\includegraphics[width=8cm]{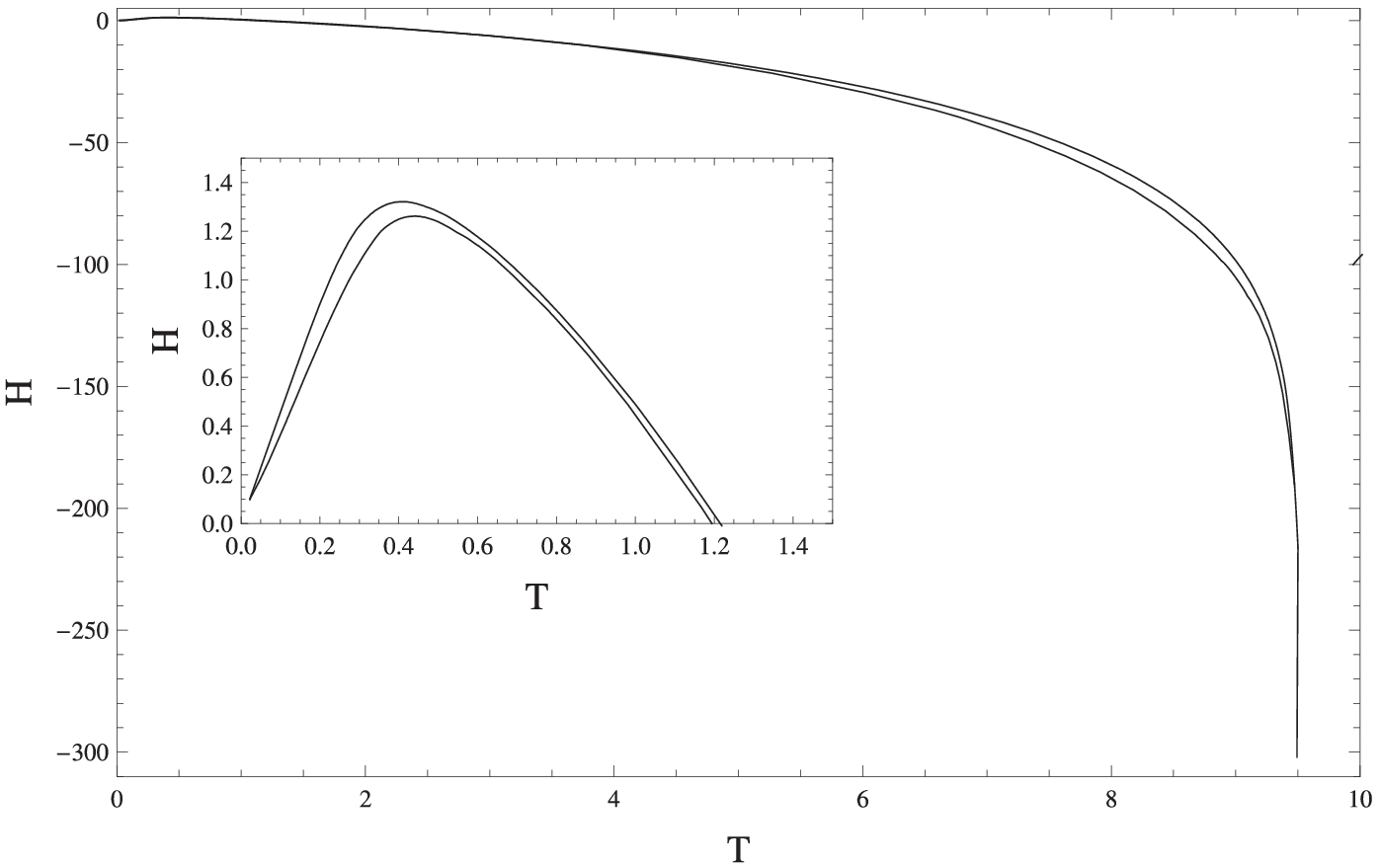}
\caption { \small {Arnold tongue with winding number
$\mathrm{w}=1/3$ for Potts model for $Q=1.1$, $J=-1$, $\gamma=3$ (the insert shows details in the area
$H>0$).}} \label{3nor}
\end{center}
\end{figurehere}
\begin{center}
\begin{figure*}
\begin{tabular}{ c c }
\small{(a)} & \small{(b)}\\
\begin{figurehere}
\includegraphics[width=7cm]{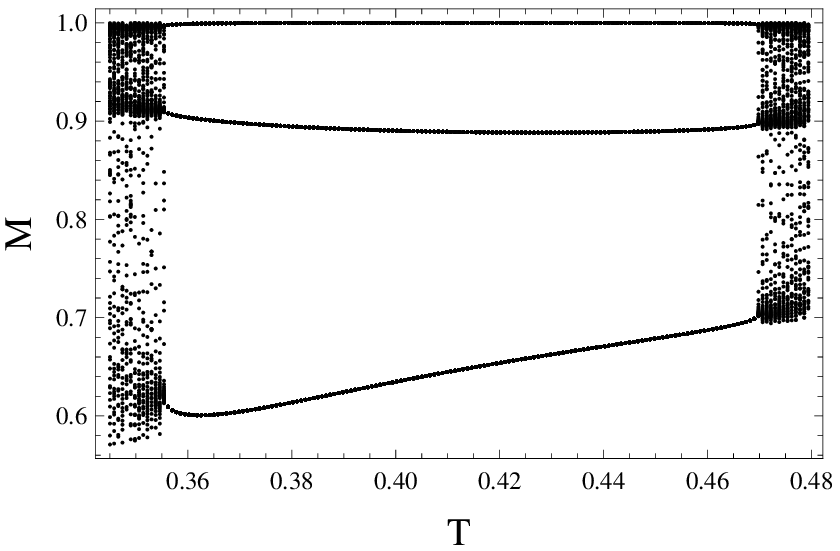}
\end{figurehere} &  \begin{figurehere}
\includegraphics[width=7cm]{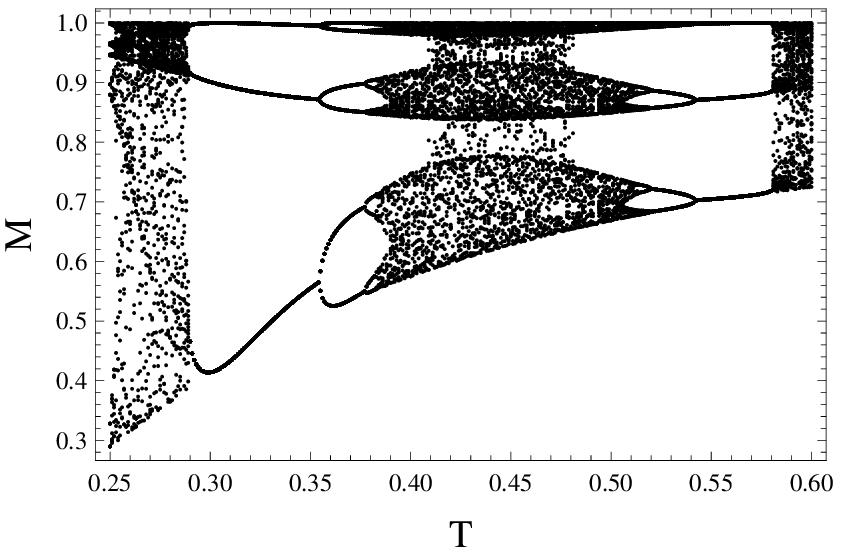}
\end{figurehere}\\
\end{tabular}
\caption {\small{ The plot of $M$ (magnetization) versus
temperature $T$ in the three periodic window for $Q=1.1$, $J=-1$, $\gamma=3$ (a) $H=1.3$; (b) $H=1.2$. \label{figm3}}}
\end{figure*}
\end{center}

Curves found from (\ref{50}) and (\ref{51}) form Arnold
tongue with winding number $\mathrm{w}=1/3$ (Fig.~\ref{3nor}). 
It corresponds to  the area of existence of the three - periodic
modulated phase (3M0, \textit{i.e.} $3\ast2^0=3$ periodic).

One can see from Fig.~\ref{3nor} that in the area $H>0$ the
mapping presents interesting behavior. Firstly, when the line $H$
intersects only the upper curve of the Arnold tongue, two edges of
the window are plainly distinguishable: the saddle-node bifurcation
takes place on both edges and the window is presented with only $3$-periodic cycle
[Fig.~\ref{figm3}(a)]. 

\begin{center}
\begin{figure*}
\begin{tabular}{ c c }
\small{(a)} & \small{(b)}\\
\begin{figurehere}
\includegraphics[width=7cm]{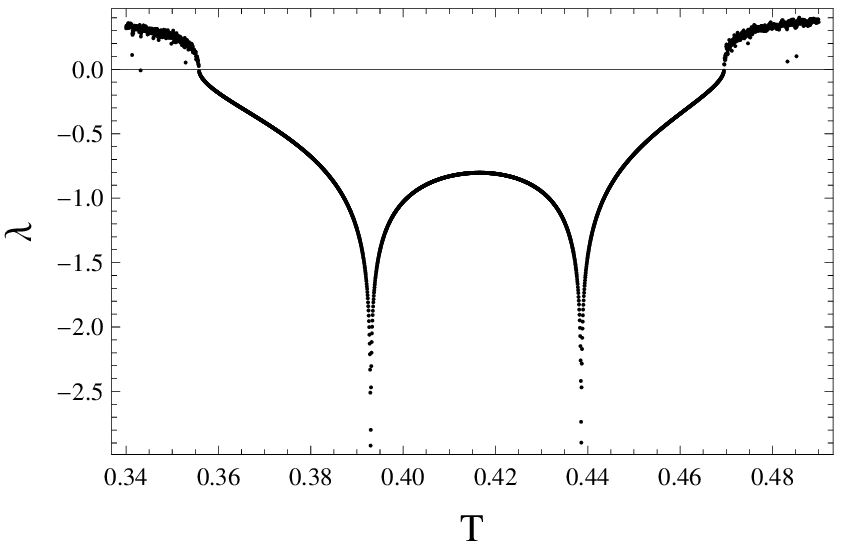}
\end{figurehere} &  \begin{figurehere}
\includegraphics[width=7cm]{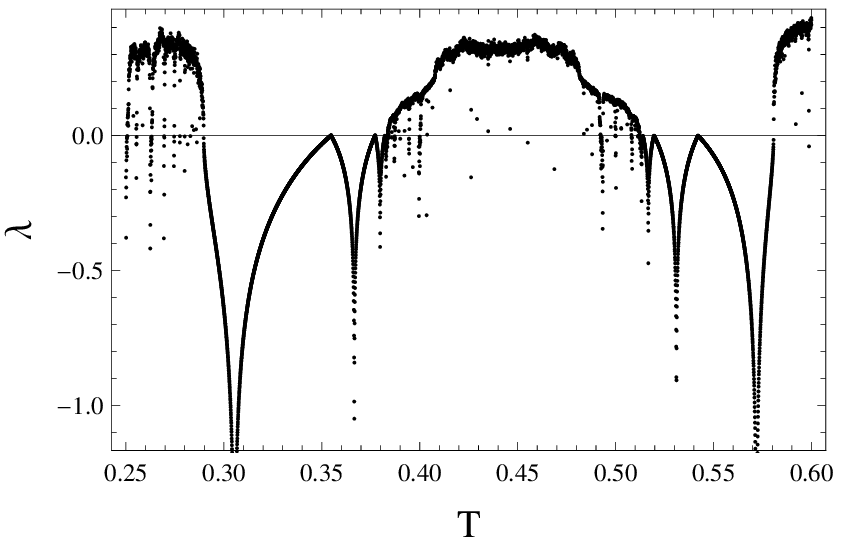}
\end{figurehere}\\
\end{tabular}
\caption {\small{  The plot of the Lyapunov exponent for the
Potts-Bethe mapping [Eq.~(\ref{7})] in the three periodic window versus temperature $T$ for the case $Q=1.1$, $J=-1$, $\gamma=3$ (a) $H=1.3$;   (b) $H=1.2$.  \label{figl3}}}
\end{figure*}
\end{center}

Transition between chaotic state and three - periodic
modulated phase (3M0) occurs.
Secondly when the line $H$ intersects lower line of the Arnold tongue, $3\ast2^1=6$ periodic cycle
appears in a form of a bubble, which corresponds to six - periodic modulated phase (3M1, \textit{i.e.} $3\ast2^1=6$ periodic). This indicates 3M0-3M1 transition. It is obvious that if we continue lowering the values of $H$ new
bubbles will appear and ultimately the chaotic region in the
window will be reached [Fig.~\ref{figm3}(b)]. However, for $H<0$
the saddle-node bifurcation occurs only at one edge of the window.
For thorough illustration of the effect, we also
present the plots of Lyapunov $\lambda$ exponents in Fig.~\ref{figl3}.

We also investigated the behavior of Feigenbaum $\alpha$ and $\delta$ exponents for three periodic window. In comparison with the period doubling, the values of $R_n$ will be found from the following condition:
\begin{equation}
f^{(3\ast2^{n})}_{R_n}(x^{*})=x^{*}.\label{53}
\end{equation}

Here again $\alpha$ and $\delta$ converges to the values in
Eqs.~(\ref{25b}) and (\ref{25a}). For example in the case $J=-1$, $T=1$, $\gamma=3$ and $Q=0.8$
for $n=10$, $\delta=4.669160924$ and $\alpha=2.502899839$.  For the $const'$ in Eq.~(\ref{26.1})
for the same values of parameters we obtained the following value: $const'=-0.0167...$ These
results confirm once more the universality hypothesis for rational mapping, which describes $Q$-state Potts model on Bethe lattice.

\section{Arnold Tongues \\through Feigenbaum \\ Exponents and Critical \\ Temperatures \label{arnold1}}

Constructing Arnold tongues for rational
mappings like Eq.~(\ref{7}), which describe real statistical
systems, is a very complicated and laborious procedure.  But it
turns out that universality hypothesis (see Sec.~\ref{feig}) is in
deep relation with this problem.
 It can help in finding a simpler
way for constructing Arnold \\ tongues, knowing only one line in
parameter space, on which some $k$th bifurcation takes place.
Here $r_m$ will be the value of parameter [see Eq.~(\ref{32})] at $m$th
bifurcation point. From the scaling relation for $r_m$ and Eq.~(\ref{31.2}) we can obtain $r_m$ in terms of
$r_k$ ($k \leq m$), $R_{\infty}$ and $\delta$:
\begin{equation}
r_m=\frac{\left(r_k \delta ^k-R_{\infty } \delta
   ^k+R_{\infty } \delta ^m\right)}{\delta ^{m}}\label{39}
\end{equation}
while $R_{\infty}$ from Eq.~(\ref{26.1}):
\begin{equation}
R_{\infty}=\frac{\text{$R_n$$\delta $}_{n+1}-R_n}{\delta -1}.
\label {39.1}
\end{equation}

The convergence of the $R_{\infty}$ to its real value is dependent
on the speed of convergence of $\delta$ exponents (for $Q>1$ $n=4$
or $5$ and in the area $Q<1$ $n=5$  or $6$ is enough).

Since $r_1$ can be found analytically in contrast with bigger
$n$'s, it would be convenient to take $k=1$, and $m>1$ [we already have
$r_1$, after solving Eq.~(\ref{33})]:

\begin{subequations}
\begin{align}\label{40a}
& r_2=\frac{r_1+R_{\infty }\delta  -R_{\infty }}{\delta }, \hspace{10pt} for \hspace{5pt} m=2,\\ \label{40b}
& r_3=\frac{R_{\infty } \delta ^2+r_1-R_{\infty }}{\delta ^2},\hspace{10pt} for \hspace{5pt} m=3.
\end{align}
\end{subequations}

For obtaining Arnold tongue analog with winding numbers $\mathrm{w}=1/2$
and $\mathrm{w}=2/4$ we have to find $r_2$ and $r_3$
for different $T$'s and $H$'s. The problem is that Eq.~(\ref{23}) is true for big $n$'s. However, Eq.~(\ref{23})
remains true if we replace $const$ with $c_n$. In this case
\begin{subequations}
\begin{align}\label{41a}
& r_2=\frac{\frac{c_2 \left(r_1-R_{\infty }\right)}{c_1}+ R_{\infty
   }\delta }{\delta },
  \hspace{10pt} for \hspace{5pt} m=2,\\ \label{41b}
&r_3=\frac{R_{\infty } \delta ^2+\frac{c_3 \left(r_1-R_{\infty
   }\right)}{c_1}}{\delta ^2},
 \hspace{10pt} for \hspace{5pt} m=3.
\end{align}
\end{subequations}
As one can see, the value of $r_2$ and $r_3$ from Eqs.~(\ref{40a}) and (\ref{40b}) are as closer to its real value in Eqs.~(\ref{41a}) and (\ref{41b}) as closer $\frac{c_1}{c_2}$ and $\frac{c_1}{c_3}$ is to $1$. Values of $\frac{c_1}{c_2}$ e.g. for investigated in the previous section $Q=0.8$ and $Q=1.2$ are $\frac{c_1}{c_2}=1.46087$ and $\frac{c_1}{c_2}=1.93921$; $\frac{c_1}{c_3}=1.48275$ and $\frac{c_1}{c_3}=2.41653$ respectively (for $T=1$). Hence Eqs.~(\ref{40a}) and (\ref{40b}) can be regarded as an approximate formula for constructing Arnold tongues.

The faster is convergence of Feigenbaum $\delta$
exponent, the closer $\frac{c_1}{c_2}$
and $\frac{c_1}{c_3}$ are to $1$. Hence one can see that there is
another reason for investigating the behavior of the Feigenbaum
exponents.

We can see from Figs.~\ref{ar1} and \ref{ar2}, that for the cases $Q=1.2$ and $Q=0.8$ there are three critical temperatures $T_n$, $T_{c_1}$ and $T_{c_2}$, at
which in absence of external magnetic field transition between
different phases takes place:

\noindent 1. $T_n$ is the temperature corresponding to
paramagnetic - anti-ferro\-mag\-ne\-tic phase (P-2AF1) transition (known as Neel
temperature). The dependence on $Q$ is shown in Fig.~\ref{fig5}(a);

\noindent 2. $T_{c_1}$ is the one corresponding to
anti-ferro\-mag\-ne\-tic - four - periodic modulated phase (2AF1-2M2)
transition [Fig.~\ref{fig5}(b)];

\noindent 3. $T_{c_2}$, the temperature at which
transition between four - periodic and eight - periodic modulated phases (2M2-2M3) occurs (the $Q$ dependence of this temperature qualitatively is the same as of $T_{c_1}$).

Investigating the behavior of the Lyapunov exponent we found that critical temperatures $T_{c_1}$  and $T_{c_2}$  exist not for all $Q$ in the area $1<Q<2$ (the fact that Lyapunov exponent does not become zero indicates the absence of bifurcation, hence the second order phase transition). After $Q=1.5681$ there is no more critical temperature like $T_{c_1}$ and after $Q=1.5115$ there is no more critical temperature like $T_{c_2}$.

\begin{center}
\begin{figurehere}
\begin{tabular}{ c }
\small{(a)} \\ 
\begin{figurehere}
\includegraphics[width=7cm]{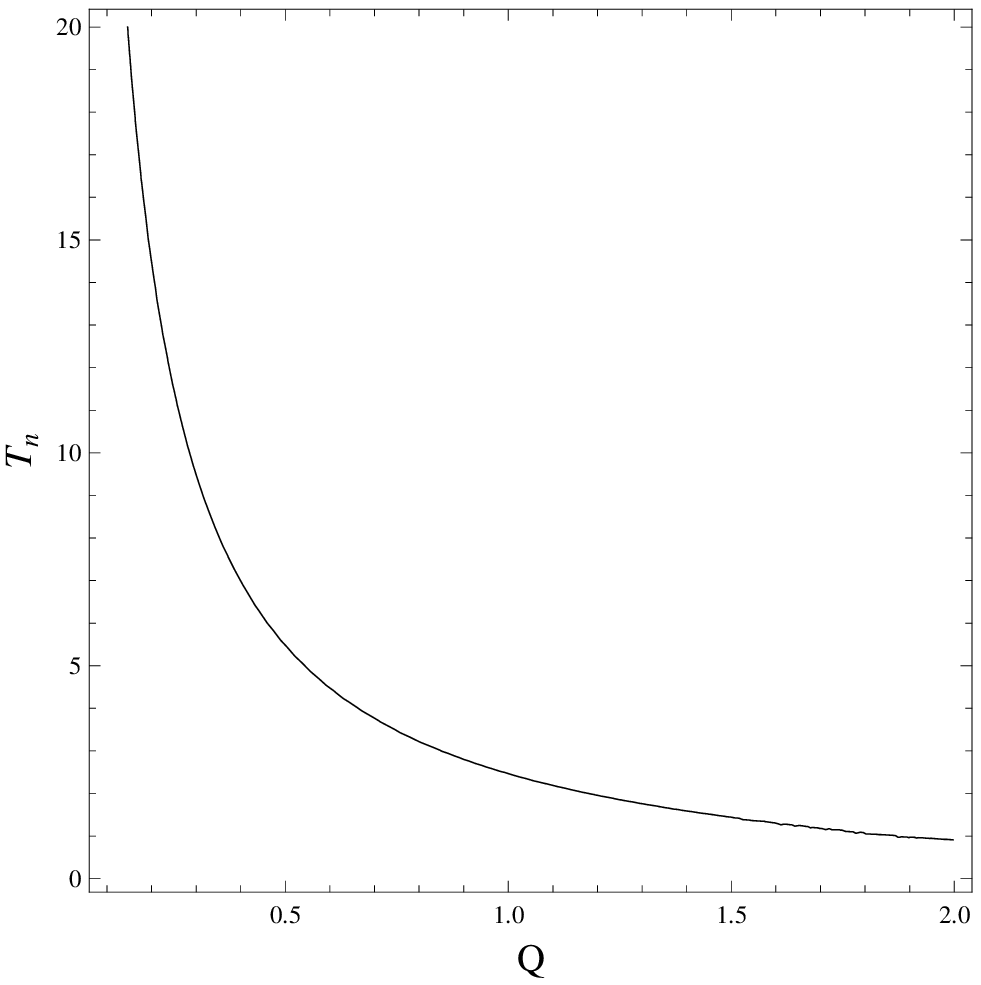}
\end{figurehere}\\ \small{(b)}\\  \begin{figurehere}
\includegraphics[width=7cm]{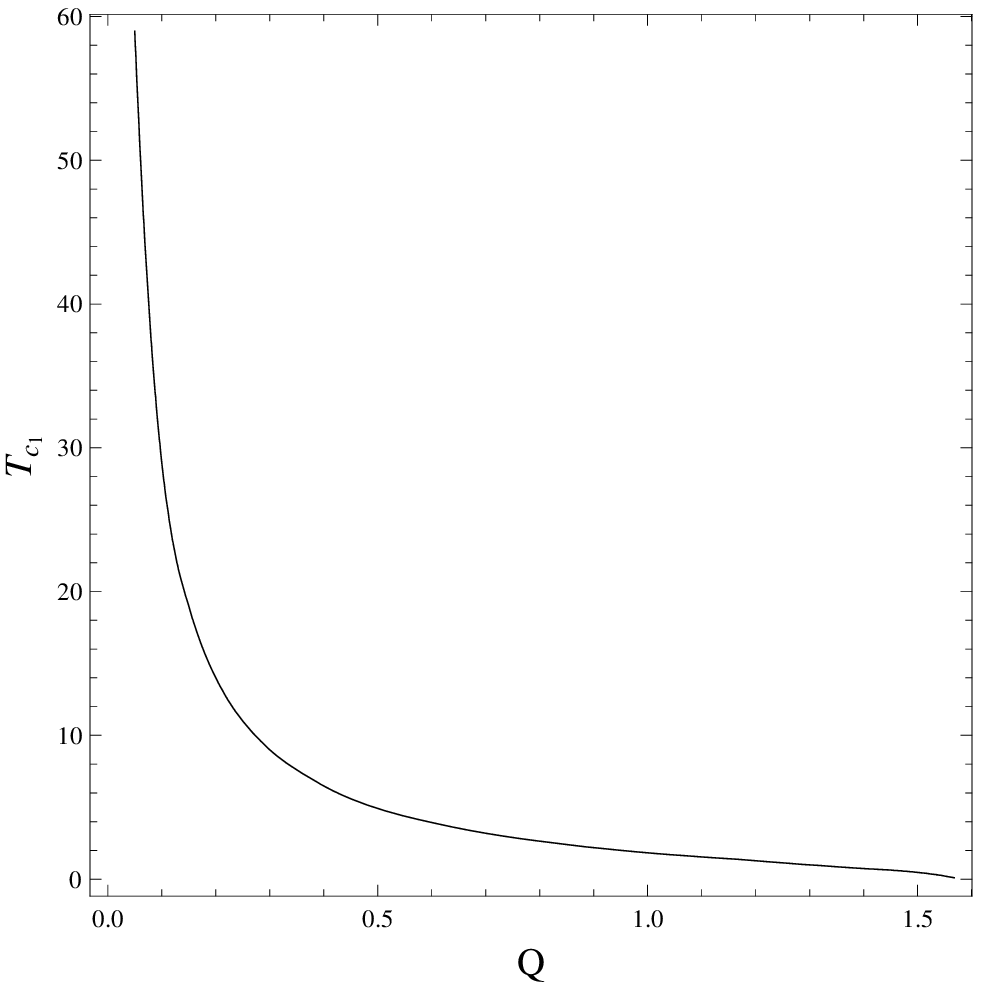}
\end{figurehere}\\
\end{tabular}
\caption {\small{  The plot of (a) $T_n$ (Neel temperature) versus $Q$ for $J=-1$, $\gamma=3$; (b) $T_{c_1}$ (critical temperature corresponding to transition between antiferromagnetic and four periodic modulated phases in absence of external magnetic field) versus $Q$ for $J=-1$, $\gamma=3$.  \label{fig5}}}
\end{figurehere}
\end{center}

Figure~\ref{ar1} shows that there is some $T_c$ (upper bound
temperature), below which the system undergoes P-2AF1-P phase
transition between paramagnetic and anti-ferromagnetic phases, \\
which exists only in the area $Q>1$ (e.g. for $Q=1.2$, $J=-1$, $\gamma=3$ $T_c=5.2606340$).

\section{Conclusion \label{concl}}

In this paper we investigated $Q$-state Potts model in an
external magnetic field on the Bethe lattice for non-integer $Q$. The model was exactly
solved by means of recurrence relation technique and a one-dimensional
rational mapping was obtained.

We pointed the relation between phase transition of
second order and bifurcation points. The phase structure of the
model is investigated by constructing Arnold tongues. For given
statistical model they correspond to the area of existence of
different phases, \textit{i.e.} to phase diagrams. We have obtained analogs of
Ar\-nold tongues with winding numbers $\mathrm{w}=1/2$ and $\mathrm{w}=2/4$
in period doubling regime. We found the transitions P-2AF1,
P-2AF1-2M2-2AF1,  P-2AF1-2M2-2M3-2AF1 in the case of fixed $H$
and P-2AF1-P at fixed $T$. The behavior of the mapping describing the system is
sensitive to the values of $Q$: case $0<Q<1$ in contrast to
$1<Q<2$. We also observed the Lyapunov exponent that describes
the large variety of phase transitions that occur in the Potts
model. We compared chaotic regions varying temperature and
magnetic field and pointed that in the first case it was richer.

We studied the convergence of the Fei\-gen\-baum $\alpha$ and
$\delta$ exponents for the above mentioned rational mapping for
$0<Q<1$ and $1<Q<2$ separately. It was shown that for a real
statistical system these exponents coincide with the famous
universal ones with high accuracy.

We constructed the Arnold tongue analogs with winding number
$\mathrm{w}=1/3$ in the three periodic window and calculated the
$\alpha$ and $\delta$ exponents for this area. The obtained values
justified the universality hypothesis once more.

We have found an approximate method for constructing Arnold
tongues through Feigenbaum exponents, based on universality
hypothesis. The main advantage of this kind of approach is the simplicity of numerical
calculations in contrast to the rigorous treatment of the problem.

\section{Acknowledgement \label{Ack}}

The authors are grateful to V. V. Hovhannisyan for
useful discussions.

This work was partly supported by 1518-PS, PS-1981, PS-2033 ANSEF and
ECSP-09-08-sasp NFSAT research grants.

\onecolumn
\hspace{50pt}

\begin{table*}
\begin{center}
\caption[Table]{Feigenbaum exponents for Potts-Bethe mapping [Eq.~(\ref{7})] for $T=1$; $J=-1$; $\gamma=3$. \label{table}}
\begin{tabular}[b]{ c c c c c }
               \hline
               \hline
               &period &  &  &\\
               $Q$ &doubling  &  $R_n$ & $\alpha$ & $\delta_n$   \\ \cline{1-5}
               & & & & \\
               &$2^1=2$   & 6.148585490... & & \\
               &$2^2=4$   & 5.624678677... & 3.068429829...& 5.321086266... \\
               &$2^3=8$   & 5.526220056... & 2.580752880...& 4.731522199... \\
               &$2^4=16$  & 5.505410977... & 2.519238533...& 4.688360097... \\
               0.8 &$2^5=32$  & 5.500972521... & 2.506269470...& 4.672585925... \\
               &$2^6=64$  & 5.500022629... & 2.503627065...& 4.670011847... \\
               &$2^7=128$ & 5.499819226... & 2.503059457...& 4.669364508... \\
               &$2^8=256$ & 5.499775665... & 2.502940259...& 4.669237820... \\
               &$......$ & ...... & ...... & ......\\
               &$2^{\infty}=\infty$ &5.499764337...  & & \\

               \hline
               & & & & \\
               &$2^1=2$   & 3.931868660... &  &  \\
               &$2^2=4$   & 2.569022501... & 3.358835598... & 6.792470680... \\
               &$2^3=8$   & 2.368381788... & 2.654374001... & 5.057633178...\\
               &$2^4=16$  & 2.328710918... & 2.534475897... & 4.745030314...\\
               &$2^5=32$  & 2.320350408... & 2.509653084... & 4.685552342...\\
               1.1 &$2^6=64$  & 2.318566091... & 2.504352497... & 4.672658688...\\
               &$2^7=128$ & 2.318184228... & 2.503218367... & 4.669946004...\\
               &$2^8=256$ & 2.318102458... & 2.502974448... & 4.669360473...\\
               &$2^9=512$ & 2.318084946... & 2.502922160... & 4.669235722...\\
               &$......$ & ...... & ...... & ......\\
               &$2^{\infty}=\infty$ & 2.318080392... &  &  \\
               \hline
               \hline
\end{tabular}
\end{center}
\end{table*}

\end{document}